\documentclass[aps,prl,twocolumn,showpacs,amsmath,amssymb]{revtex4-1}
\usepackage{graphicx}
\usepackage{dcolumn}
\usepackage{epstopdf}
\usepackage{multirow}
\usepackage{bm}
\begin{document}

\title{On-the-fly exact diagonalization solver for quantum electronic models}
\author{I.V. Kashin$^{1}$, V.V. Mazurenko$^{1}$}
\affiliation{
$^{1}$Theoretical Physics and Applied Mathematics Department, Ural Federal University, Mira Str.19,  620002 Ekaterinburg, Russia}
\date{\today}

\begin{abstract}
We propose a distinct numerical approach to effectively solve the problem of partial diagonalization of the super-large-scale quantum electronic Hamiltonian matrices.  The key ingredients of our scheme are the new method for arranging the basis vectors in the computer's RAM and the algorithm allowing not to store a matrix in RAM, but to regenerate it on-the-fly during diagonalization procedure. This scheme was implemented in the program, solving the Anderson impurity model in the framework of dynamical mean-field theory (DMFT). The DMFT equations for electronic Hamiltonian with 18 effective orbitals that corresponds to the matrix with the dimension of $2.4 \times 10^{9}$ were solved on the distributed memory computational cluster. 
\end{abstract}

\maketitle

Nowadays one can barely imagine a modern physical or computational problem which could be solved without application of the matrix calculus. Indeed, if any more or less intricate equation system is under our investigation [1-3], we need a reliable algorithm to treat the basic matrix operations like diagonalization. This procedure is known to be strictly necessary in the various fields of computational science from the latent semantic analysis in linguistics [4] to the investigation of correlated crystals in condensed matter physics [5]. At present the modern numerical algorithms and libraries such as LAPACK [6] can be used for the full diagonalization of the matrices with dimension of a few 10000. But as the size of the matrix to be dealt with grows further, the natural solution of the problem is to replace the full diagonalization by partial one. This approach is applicable if the key information about physical system can be extracted from a small part of the eigenvalue spectrum.

That is the case of the modern quantum physics. In order to describe a physical system at low temperatures we need to calculate only little amount of the lowest eigenvalues. One uses them to model the ground state and the spectrum of low-energy excitations. And this is exactly what the partial diagonalization technique is designed to.

In our study we use the exact diagonalization (ED) approach to solve the Anderson impurity model [7]. In terms of this model the impurity is assumed to be embedded in an effective electron bath instead of the being placed into crystal lattice. It allows us to investigate the behaviour of the system in the strongly correlated regime, when the electron-electron interaction and the kinetic energy are of the same order. Along with the natural physical application, the same model is exploited in the self-consistent cycle to solve the equations of dynamical mean-field theory (DMFT) [7].

In the simplest case the Hamiltonian of the Anderson model can be written in the following form:

\begin{eqnarray}
H = \sum_{\sigma} E_{0} d^{+}_{0 \sigma} d_{0 \sigma} + U n_{0 \uparrow} n_{0 \downarrow} \\
+ \sum_{p,\sigma} [V_{p 0} c_{p \sigma}^{+} d_{0 \sigma} + V_{p 0}^{*} d_{0 \sigma}^{+} c_{p \sigma}] 
+ \sum_{p,\sigma} \epsilon_p c_{p \sigma}^{+} c_{p \sigma} , \nonumber
\end{eqnarray}
where $\sigma$ is the spin index, $c_{p \sigma}(c_{p \sigma}^{+})$ and $d_{0 \sigma}(d_{0 \sigma}^{+})$ are electron annihilation(creation) operators for the bath sites and the impurity site correspondingly, $V_{p 0}$ is the impurity-bath hybridization, $n_{0 \sigma} = d^{+}_{0 \sigma} d_{0 \sigma}$ is the particle number operator and $U$ is the on-site Coulomb interaction. 

In general case the number of bath sites is infinite. In the framework of the ED method Eq.(1) is approximated by its reduced version, constructed with the finite number of bath sites. Appropriateness of this approximation is thoroughly discussed in Ref.[7]. 

The sum of the bath and impurity sites forms the total number of sites, i.e. the number of the effective electronic orbitals $N_{s}$ taken into account. In the pioneer study [7] $N_{s}$ was ranged from 5 to 12, but using power of the modern computational clusters it can be increased up to 17 [8]. This limit generally rises because of restriction of the computation resourse, typically being the amount of RAM per each processor. Therefore to make a step forward we need to find a way to modernize the algorithms in use to reduce the RAM space required. The natural solution is to develop an algorithm allowing not to keep the matrix in RAM, but to recalculate it (on-the-fly) on the each step of iterative diagonalizing process. It was applied to treat the ultra-large sparse spin Hamiltonian matrix with the dimension $2.7 \times 10^{11}$ [9] and in this study we accomplished it to solve the electronic problem of the scale $2.4 \times 10^{9}$.

As one can see from Eq.(1), there are no operators changing the total number of electrons in the system. Therefore, being presented in matrix form, Hamiltonian (1) has a block structure, where each block corresponds to a fixed amount of the spin-up and spin-down electrons, and does not mix to any other one. Therefore, they can be diagonalized independently. The $N_{\uparrow}$=1 and $N_{\downarrow}$=1 block for the simple system, consisting of one impurity and one bath site is shown in Fig. 1.

\begin{figure}[t]
\includegraphics[width=0.3\textwidth,angle=0]{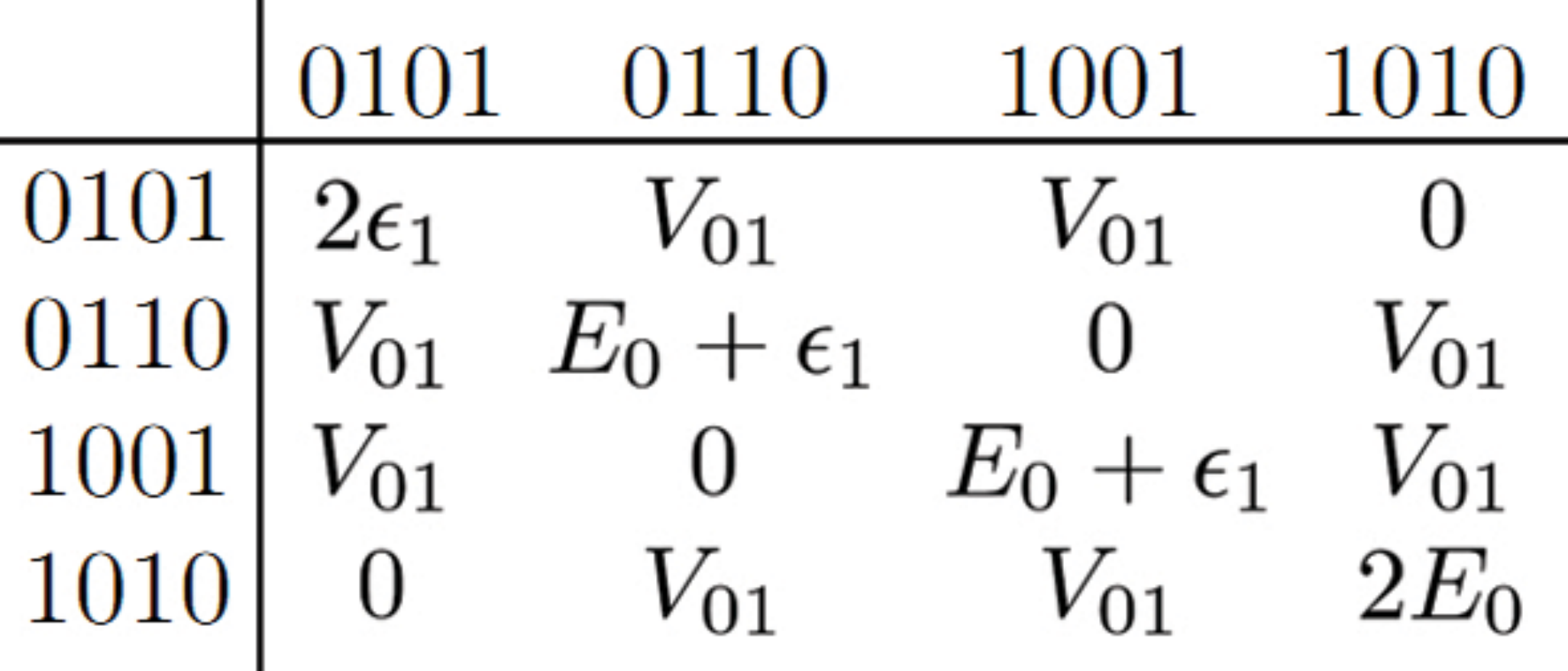}
\caption{The example of the Hamiltonian matrix block in case of $N_{\uparrow}$=1 and $N_{\downarrow}$=1 for the one impurity and one bath site system. The basis vectors are constructed in accordance with Fig. 2.}
\label{matrixblock}
\end{figure}

The main procedure of the numerical diagonalization is the matrix-vector multiplication. In this sense the most preferable approach is considered to be Arnoldi method [10], which requires at least the initial and resulting vector to keep them in the computer RAM. If we generate the electronic Hamiltonian instead of its storage in memory, we have to make it in an effective way. In this paper we propose a high-performance method for the partial diagonalization of the electronic Hamiltonian matrices. It consists of the basis optimization module and matrix-vector multiplication module. Each of them we will describe below.

{\it Basis optimization.}
The first step to diagonalize Anderson-type electronic Hamiltionan is to optimize the way of the basis treating. Mainly we need a fast algorithm to handle the basis states. Each of the states can be determined by combination of the occupation numbers (Fig. 2), which, in turns, can be represented in binary code of $2N_s$ length ($N_s$ digits per each spin projection). 

\begin{figure}[t]
\includegraphics[width=0.3\textwidth,angle=0]{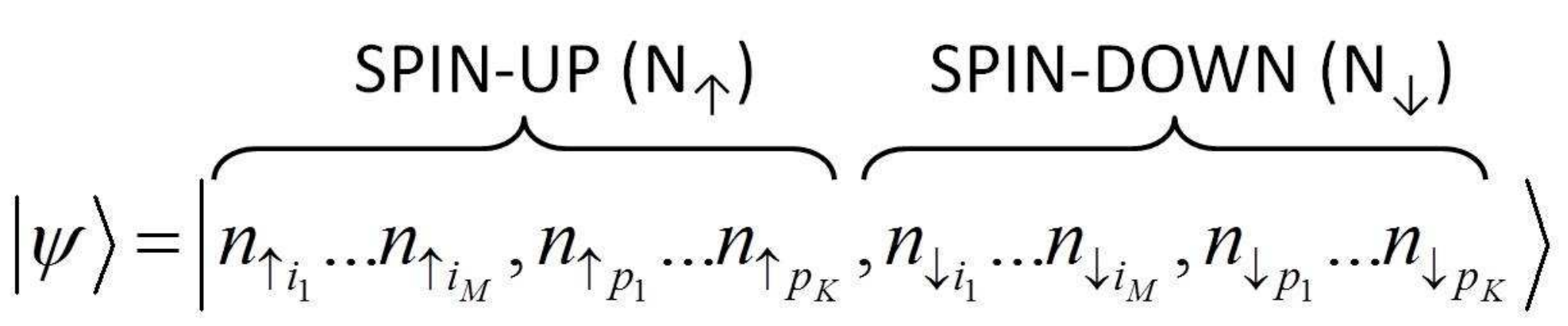}
\caption{Basis vector in occupation numbers representation. $n_{\uparrow(\downarrow) i(p)}$ – number of spin-up(down) electrons on the impurity(bath) states, $N_{\uparrow(\downarrow)}$ – total number of spin-up(down) electrons in corresponding configuration.}
\label{basisvector}
\end{figure}  

In this representation, the full basis for the quantum system, consisting of 17 effective electronic orbitals, is a huge dimension of $2*17*4^{17}$  binary bits, which requires about 6 Tb RAM. If we keep this basis in the block form, the amount of the required memory will be decreased to 2.5 Gb, but it is still not suitable. 

We propose a new method of the basis storage. In Table II (col. 1) we show the basis of the Hamiltonian matrix block $N_{\uparrow}=2, N_{\downarrow}=1$ for the simple quantum system of three sites. It is shown that if the corresponding numbers in the binary representation are allocated in ascending order, both (spin-up and spin-down) electronic configurations also were built in the same way. One can see that spin-up binary representation changes only when the spin-down one overflows like a scale of notation.

In general case, if we construct all the basis vectors, corresponding to any block $N_{\uparrow}, N_{\downarrow}$ for the system with arbitrary $N_s$ the spin-up configuration being external  is observed going from the lowest to the highest binary representation only once, whereas the spin-down one being internal goes $C^{N_s}_{N_{\downarrow}}$ times. Thereby, to build all basis vectors we should treat all matrix blocks $N_{\uparrow}, N_{\downarrow}$ successively, where $N_{\uparrow, \downarrow}$ ranges from 0 to $N_s$.

We stress that for the system of $N_s$ electronic orbitals there is no difference between spin-up and spin-down electronic configurations, and they are completely determined only by number of electrons $N_{\uparrow}$ and $N_{\downarrow}$. Therefore, to construct any basis vector of this system one needs to store just all possible sets of electronic configurations, characterized by the fixed number of electrons, varying from 0 to $N_s$, regardless to the spin projection. It is illustrated in the Table II (col. 2-3), where each basis vector of the matrix block was transformed into combination of the corresponding binary representations. In terms of this example we have only 8 configurations with the length of 3 ($N_s$) binary digits (Table I) to produce any basis state of the whole Hamiltonian matrix instead of 9 vectors with the length of 6 ($2N_s$) binary digits to treat only single block $N_{\uparrow}=2, N_{\downarrow}=1$ not to mention the full basis, containing 64 vectors. Hence, in that case we have to store $3 \times 8=24$ binary digits instead of $6 \times 64=384$ ones. In this way, required RAM space for reproducing any basis state in binary representation for quantum system with $N_s = 17$ is 1.7 Mb. It is of crucial importance when we are dealing with ultra-large-scale matrices. 

\renewcommand{\arraystretch}{1.4}
\renewcommand{\tabcolsep}{0.6cm}
\begin{table}[h!] 
\caption{All possible electronic configurations for the quantum system with $N_s$ = 3. $n_{conf}$ is the ordinal number of configuration.} 
\begin{tabular}{|c|c|c|c|c|}
\hline \hline
\multirow{2}{*}{\large $\mathbf{n_{conf}}$} & \multicolumn{4}{c|}{\textbf{Number of electrons}} \\
\cline{2-5}
& \textbf{0} & \textbf{1} & \textbf{2} & \textbf{3} \\
\hline
\textbf{\textit{1}} & 000 & 001 & 011 & 111 \\
\hline
\textbf{\textit{2}} & - & 010 & 101 & - \\
\hline
\textbf{\textit{3}} & - & 100 & 110 & - \\
\hline \hline
\end{tabular}
\end{table}

\renewcommand{\arraystretch}{1.4}
\renewcommand{\tabcolsep}{0.2cm}
\begin{table}[h]
\caption{Basis and its representation in terms of individual electronic configurations for the quantum system with $N_s$ = 3. $n_{conf}$ is the ordinal number of the configuration.} 
\begin{tabular}{|c|c|c|}
\hline \hline
\multirow{3}{*}{\textbf{Basis vector}} & \multicolumn{2}{c|}{\textbf{Electronic configurations}} \\
\cline{2-3}
& Spin up, $N_{\uparrow}=2$, & Spin down, $N_{\downarrow}=1$, \\
& $n_{conf} / Conf$ & $n_{conf} / Conf$ \\
\hline
 011 001 & \textbf{\textit{1}} / 011 & \textbf{\textit{1}} / 001 \\
\hline
011 010 & \textbf{\textit{1}} / 011 & \textbf{\textit{2}} / 010 \\
\hline
011 100 &\textbf{\textit{1}} / 011 & \textbf{\textit{3}} / 100 \\
\hline
101 001 & \textbf{\textit{2}} / 101 & \textbf{\textit{1}} / 001 \\
\hline
101 010 & \textbf{\textit{2}} / 101 & \textbf{\textit{2}} / 010 \\
\hline
101 100 & \textbf{\textit{2}} / 101 & \textbf{\textit{3}} / 100 \\
\hline
110 001 & \textbf{\textit{3}} / 110 & \textbf{\textit{1}} / 001 \\
\hline
110 010 & \textbf{\textit{3}} / 110 & \textbf{\textit{2}} / 010 \\
\hline
110 100 & \textbf{\textit{3}} / 110 & \textbf{\textit{3}} / 100 \\
\hline \hline
\end{tabular}
\end{table}

{\it Matrix-vector multiplication module.} 
One of the most essential aspects of the numerical on-the-fly diagonalization is the optimization of the matrix-vector multiplication. The author of Ref. 8 proposed an algorithm, consisted in building some new individual vector on each processor from only the necessary pieces of original one (Fig. 3) and use it to perform multiplication. This vector is much smaller, because of a sparse structure of the Hamiltonian matrix.

\begin{figure}[]
\includegraphics[width=0.45\textwidth,angle=0]{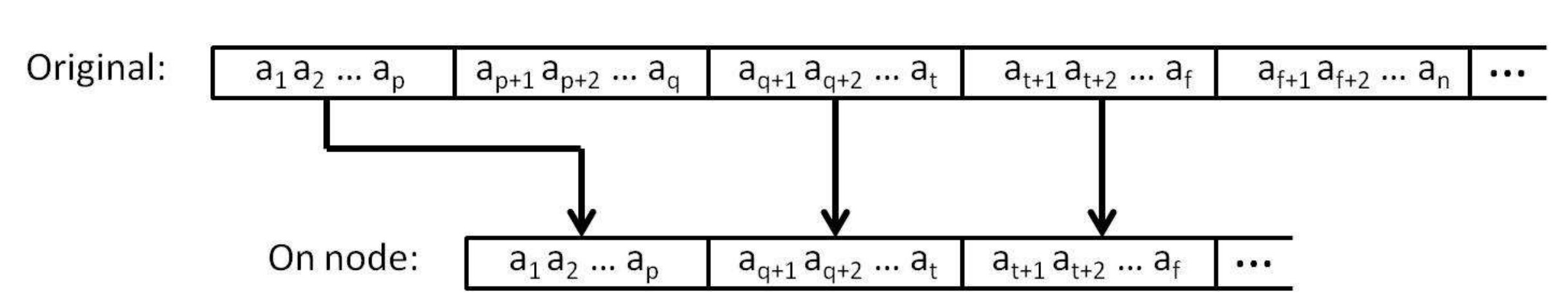}
\caption{Construction of the vector on node from the pieces of original vector.}
\label{vectorconstruct}
\end{figure}  

That is suitable to minimize the number of interprocessor communications, but in the case of a super-large-scale matrix even one piece of original vector is large, which complicates its transfer between nodes. To solve this problem we used the fact that because of sparse structure of the matrix not every element of the vector contributes to the result. And the contributing elements are distributed facultatively. Therefore, we can transfer not the whole vector pieces, but its short versions from minimum indexed to maximum indexed contributing element (Fig. 4).  It means that the very first contributing element $a_{min} >= a_1$ and the very last one $a_{max} <= a_p$ can be readily defined, making possible to cut the "left and right tails"  $[a_1, a_{min-1}]$ and $[a_{max+1}, a_p]$. 

\begin{figure}[]
\includegraphics[width=0.45\textwidth,angle=0]{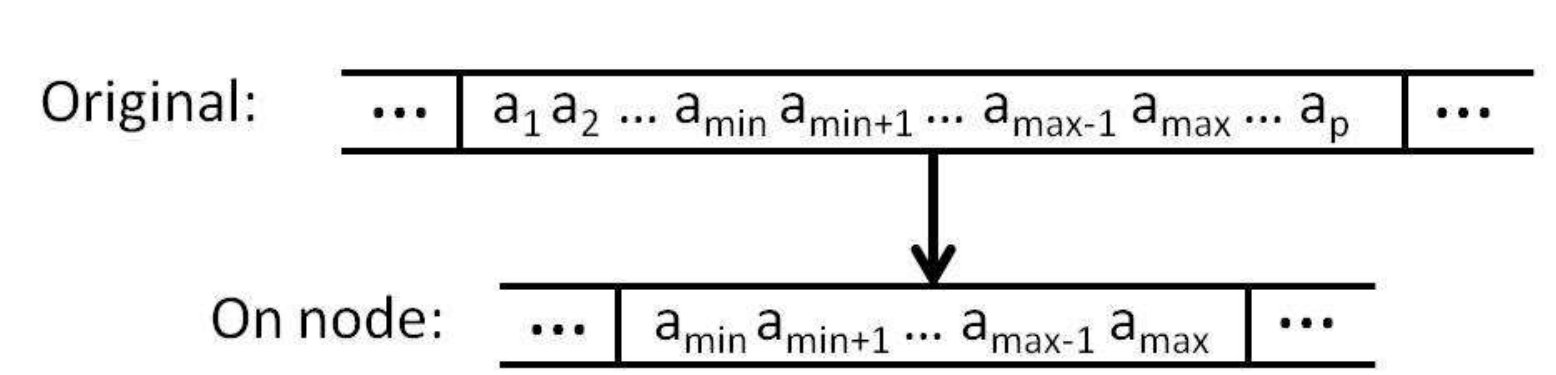}
\caption{Transfer of minimum to maximum necessary elements of the piece.}
\label{frommintomax}
\end{figure}  

For the calculations with 128 and more processors engaged the total amount of the transmittable data was reduced to 50-55 \% in comparison with whole pieces transfer, providing a productivity boost and decrease of the required RAM.

{\it Performance estimation.} To demonstrate performance the developed program was applied to solve the Hubbard model [7] on the square lattice by using the DMFT method. Such a model is widely used for the electronic and magnetic structure simulation of the superconducting copper oxides [11].

One of the most important criterion of parallel method performance is its scalability. To test it we carried out the series of partial diagonalizations of the matrix $11778624 \times 11778624$, which corresponds to the largest block of the Hamiltonian matrix of the electronic model with 14 effective orbitals. All the calculations were performed on $N_p$ processors with the clock rate of 2.2 GHz. As it is shown in Fig. 5, test curve is close to the ideal one, therefore we obtain a remarkable scaling.  Results of these tests were compared with ones, provided from program based on compressed row storage (CRS) [8] method to keep Hamiltonian matrix in RAM (Tables III and IV). Moreover, for a first time this model was solved with 18 effective electronic orbitals involved. The density of states, obtained for metallic (a) and insulating (b) case using Lanczos algorithm [12] are shown in Fig.6. 

\begin{figure}[]
\includegraphics[width=0.45\textwidth,angle=0]{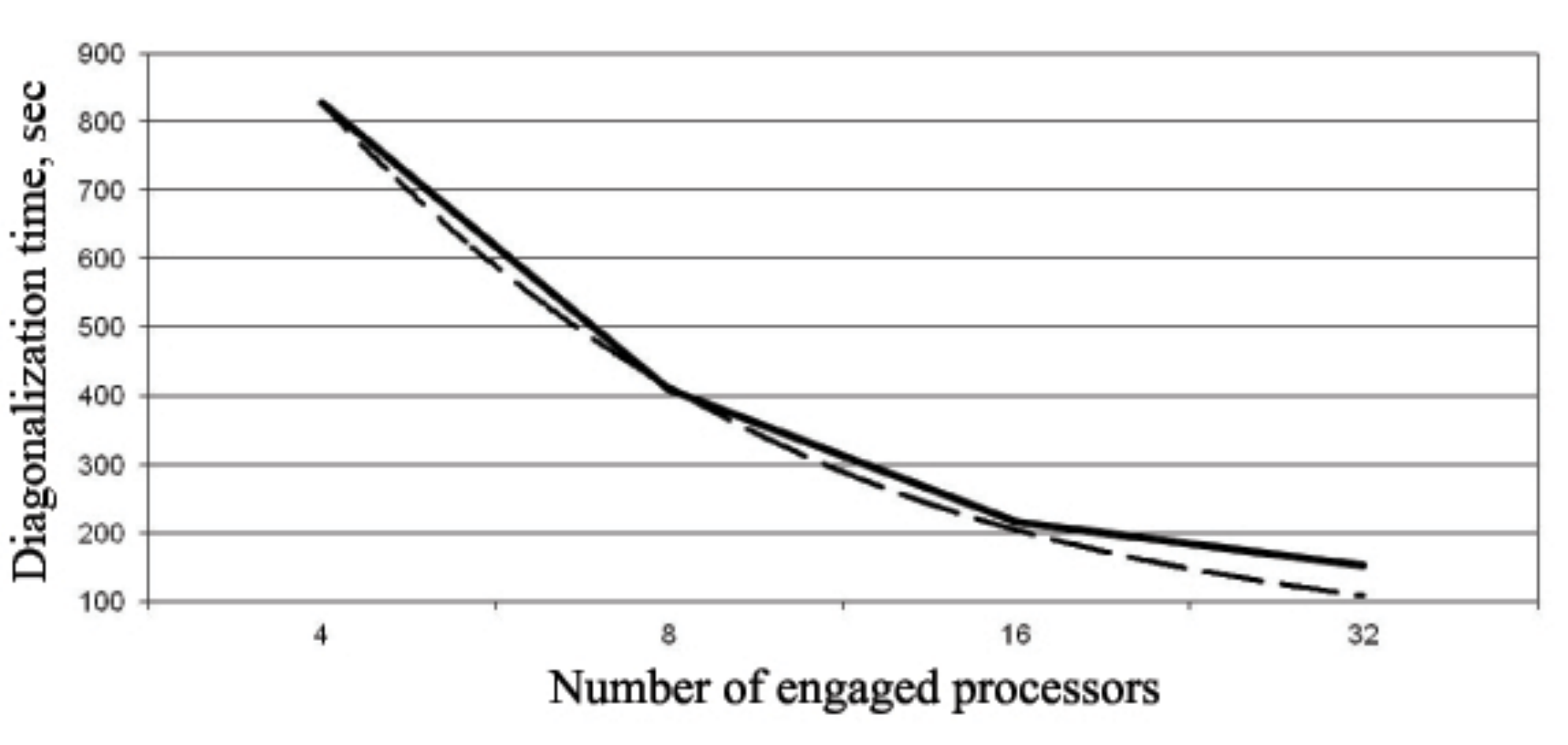}
\caption{Dependence of the diagonalization time on the number of engaged processors.}
\label{scaling}
\end{figure} 

\begin{figure}[]
\includegraphics[width=0.45\textwidth,angle=0]{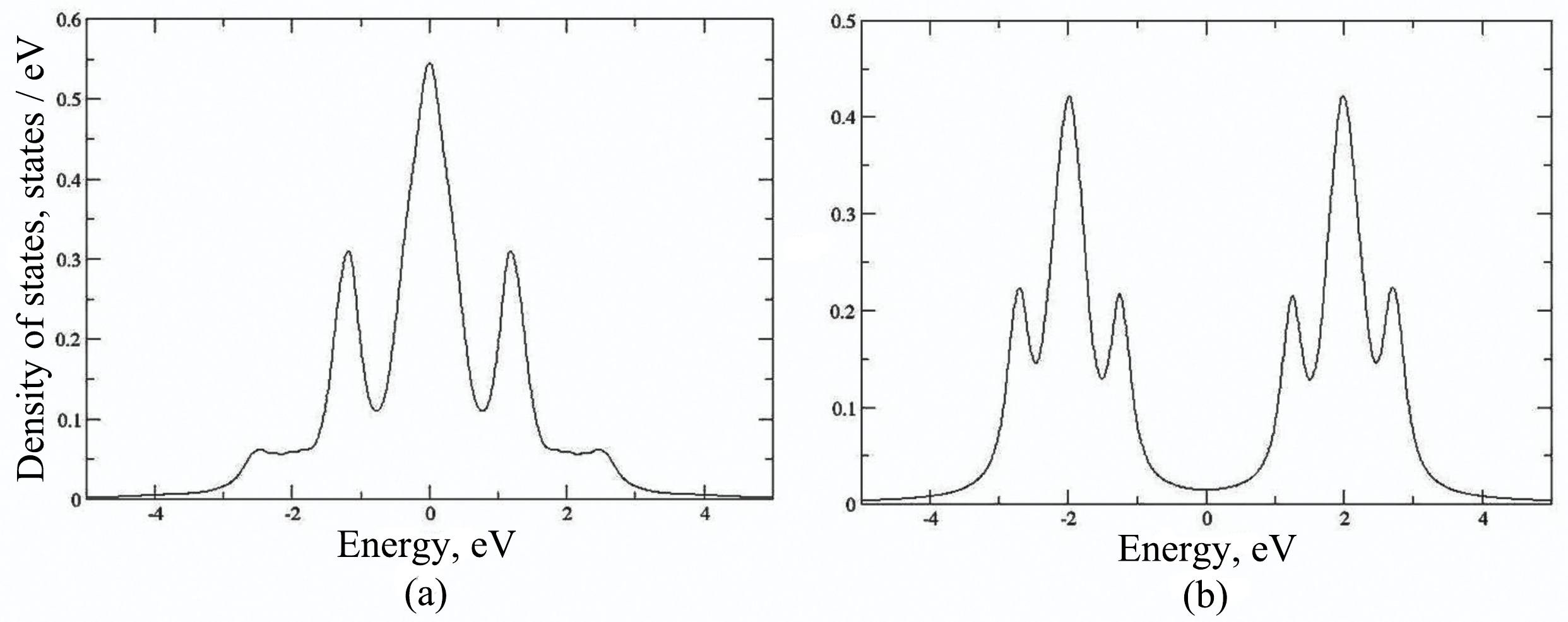}
\caption{Density of states of Hubbard model in square lattice for metallic (a) and insulator (b) cases.}
\label{dosiki}
\end{figure} 

\renewcommand{\arraystretch}{1.4}
\renewcommand{\tabcolsep}{0.12cm}
\begin{table}[h]
\caption{Performance comparison of the developed program and another one, where CRS method is used to store the Hamiltonian matrix in RAM: Memory required. } 
\begin{tabular}{|c|c|c|c|c|c|}
\hline \hline
\multirow{2}{*}{$\mathbf {N_s}$} & \multirow{2}{*}{\textbf{Matrix size}} & \multirow{2}{*}{$\mathbf {N_p}$} & \multicolumn{2}{c|}{\textbf{RAM, Mb}} & \textbf{Economy,} \\
\cline{4-5}
 & & & \textbf{On-the-fly} & \textbf{CRS} & \textbf{\%} \\
\hline
14 & 11 778 624 & 32 & 39 & 300 & 87  \\
\hline
16 & 165 636 900 & 256 & 73 & 500 & 85.4  \\
\hline
17 & 590 976 100 & 512 & 131 & 1000 & 86.9  \\
\hline
18 & 2 363 904 400 & 512 & 502 & - & -  \\
\hline \hline
\end{tabular}
\end{table}

\renewcommand{\arraystretch}{1.4}
\renewcommand{\tabcolsep}{0.3cm}
\begin{table}[h]
\caption{Performance comparison of the developed program and another one, where CRS method is used to store the Hamiltonian matrix in RAM: Diagonalization time. } 
\begin{tabular}{|c|c|c|c|c|}
\hline \hline
\multirow{2}{*}{$\mathbf {N_s}$} & \multirow{2}{*}{\textbf{Matrix size}} & \multirow{2}{*}{$\mathbf {N_p}$} & \multicolumn{2}{c|}{\textbf{Total time, sec}} \\
\cline{4-5}
 & & & \textbf{On-the-fly} & \textbf{CRS} \\
\hline
14 & 11 778 624 & 32 & 120 & 188 \\
\hline
16 & 165 636 900 & 256 & 600 & 602 \\
\hline
17 & 590 976 100 & 512 & 1480 & 1300 \\
\hline
18 & 2 363 904 400 & 512 & 6548 & - \\
\hline \hline
\end{tabular}
\end{table}

{\it Conclusion.}  The new technique for treating the super-large-scale sparse Hamiltonian matrices was developed. An effective arrangement of the basis vectors and numerical scheme allowing not to keep the matrix in the RAM, but to recalculate it during the diagonalization procedure resulted in considerable economy of the calculation resources which turned to be near 87\%. It gives us an opportunity to simulate the quantum impurity systems with 18 electronic orbitals.

{\it Acknowledgements.}
This work is supported by the grant of Russian Science Foundation (project No. 14-12-00306).

\end{document}